\documentclass[
reprint,
superscriptaddress,
amsmath,amssymb,
aps,
prb,
floatfix,
]{revtex4-2}

\usepackage{graphicx}
\usepackage{bm}
\usepackage[version=4]{mhchem}
\usepackage[bookmarks=false,colorlinks]{hyperref}
\usepackage[note-name=, use-sort-key = false]{notes2bib}

\newcommand{\lnobi}{{\ce{La3Ni2O7}}}
\newcommand{\tc}{{$T_{\rm c}$}}
\newcommand{\nef}{$N(E_{\rm F})$}

\hypersetup{
    colorlinks = true,
    citecolor = blue,
    bookmarksnumbered = true,
    urlcolor = blue
}
\usepackage[]{hyperref}
\usepackage{pdfpages}

\makeatletter
\AtBeginDocument{\let\LS@rot\@undefined}
\makeatother

\begin{document}
\title{Incommensurate spin fluctuations and competing pairing symmetries in \ce{La3Ni2O7}}

\author{Han-Xiang Xu}
\affiliation{Beijing National Laboratory for Condensed Matter Physics, Institute of Physics, Chinese Academy of Sciences, Beijing 100190, China}

\author{Daniel Guterding}
\affiliation{Technische Hochschule Brandenburg, Magdeburger Stra{\ss}e 50, 14770 Brandenburg an der Havel, Germany}

\date{\today}

\begin{abstract}
The recent discovery of superconductivity in the bilayer Ruddlesden-Popper nickelate {\lnobi} under high pressure has generated much interest in the superconducting pairing mechanism of nickelates. Despite extensive work, the superconducting pairing symmetry in {\lnobi} remains unresolved, with conflicting results even for identical methods. We argue that different superconducting states in {\lnobi} are in close competition and highly sensitive to the choice of interaction parameters as well as pressure-induced changes in the electronic structure. Our study uses a multi-orbital Hubbard model, incorporating all Ni $3d$ and O $2p$ states. We analyze the superconducting pairing mechanism of {\lnobi} within the random phase approximation and find a transition between $d$-wave and sign-changing $s$-wave pairing states as a function of pressure and interaction parameters, which is driven by spin fluctuations with different wave vectors. These spin fluctuations with incommensurate wave vectors cooperatively stabilize a superconducting order parameter with $d_{x^2-y^2}$ symmetry for realistic model parameters. Simultaneously, their competition may be responsible for the absence of magnetic order in {\lnobi}, demonstrating that magnetic frustration and superconducting pairing can arise from the same set of incommensurate spin fluctuations.
\end{abstract}

\maketitle

\section{Introduction}
The synthesis and physical properties of Ruddlesden-Popper nickelates have been studied for three decades in both experiment~\cite{zhang1994synthesis} and theory~\cite{seo1996electronic}. Shortly after the initial synthesis of \lnobi, a metal-insulator transition in oxygen-deficient samples of La$_{3}$Ni$_{2}$O$_{7-\delta}$ was found~\cite{taniguchi1995transport}. After the discovery of high-{\tc} superconducting cuprates with Cu$^{2+}$ ions ($3d^9$ configuration), interest in nickelates with Ni$^{1+}$ (in $3d^9$ configuration) also increased, as the community expected these compounds to mirror the properties of cuprates~\cite{hayward1999sodium}. Superconductivity in these materials, however, remained elusive until the discovery of superconducting thin-films of the infinite-layer nickelate Nd$_{0.8}$Sr$_{0.2}$NiO$_{2}$ with a {\tc} of around 9 to 15~K~\cite{li2019superconductivity}. In 2023, a significant breakthrough was achieved with the discovery of superconductivity in Ruddlesden-Popper phase bilayer nickelate {\lnobi} under a pressure of 14~GPa with a relatively high transition temperature of around 80~K~\cite{sun2023signatures}.

Naturally, this new high-{\tc} superconductor has since attracted much interest, while methods for its synthesis have been significantly improved~\cite{li2024pressure, wang2024bulk}. The crystal structure of {\lnobi} under pressure has meanwhile been resolved using X-ray diffraction. The material undergoes a structural phase transition from $Amam$ to $Fmmm$ (orthorhombic) and finally to $I4/mmm$ (tetragonal) as a function of pressure and temperature~\cite{Wang2024,li2024pressure}. The onset of superconductivity in the pressure range from 14~GPa to 90~GPa has been determined using resistivity measurements, which results in a phase-diagram with a superconducting dome that is very steep at the low-pressure end and falls off slowly towards higher pressures~\cite{li2024pressure}.

Several other experiments have been performed that give further insight into this fascinating class of materials. Electron energy-loss spectroscopy shows strong $p$-$d$ hybridization and underscores the importance of considering the role of oxygen $2p$ orbitals~\cite{dong2024visualization}. Experiments such as nuclear magnetic resonance spectroscopy of the $^{\rm 139}$La nuclei~\cite{dan2024spin}, muon spin relaxation~\cite{chen2024evidence}, inelastic neutron scattering~\cite{xie2024neutron} and resistance measurements~\cite{zhang2024high} indicate the presence of magnetic excitations and potential density-wave-like transitions. However, no long-range magnetic order has been found in neutron diffraction and inelastic neutron scattering experiments down to a temperature of 10~K~\cite{xie2024neutron, ling2000neutron}. In addition, angle-resolved photoemission spectroscopy has mapped the Fermi surface and the electronic band structure~\cite{yang2024orbital}, which is a helpful guide for theoretical studies.

\begin{figure*}[htb]
\includegraphics[width=\textwidth]{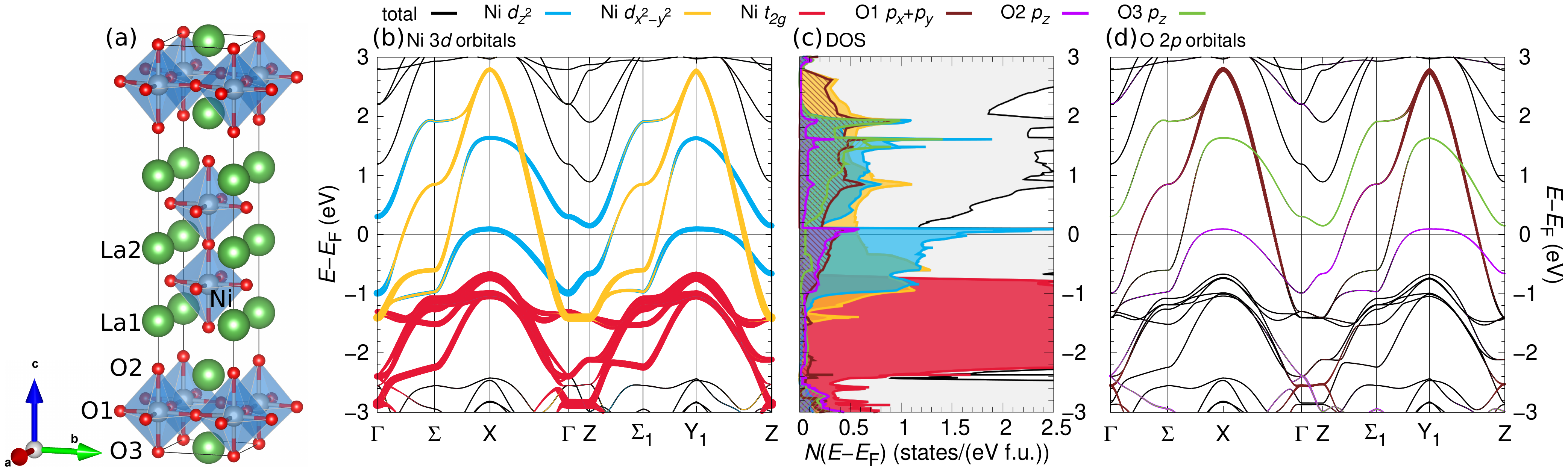}
\caption{Crystal structure, electronic band structure and density of states of {\lnobi}. (a) Crystal structure of {\lnobi} (space group $I4/mmm$, No. 139) with corner sharing \ce{NiO6} octahedra visualized in VESTA~\cite{momma2011vesta}. The labels distinguish atoms on different Wyckoff positions. O1, O2, and O3 denote the inner-layer oxygen, apical oxygen, and inter-layer oxygen respectively. (b) Band structure of {\lnobi} with orbitals weights of \ce{Ni} $3d$ orbitals. (c) Orbital-resolved electronic density of states of {\lnobi}. (d) Orbital weights of the relevant subset of \ce{O} $2p$ states. For the $k$-path we use symmetry points~\cite{setyawan2010high} $\Gamma=(0, 0, 0)$, $\Sigma=(-\eta, \eta, \eta)$, $X=(0, 0, 1/2)$, $Z=(1/2, 1/2, -1/2)$, $\Sigma_{1}=(\eta, 1-\eta, -\eta)$, $Y_{1}=(1/2, 1/2, -\zeta)$ in terms of primitive reciprocal lattice vectors, where $\eta=(1+a^{2}/c^{2})/4$, $\zeta=a^{2}/(2c^{2})$. $a$ and $c$ are lattice parameters of the conventional tetragonal cell.}
\label{fig:crystal_electronic}
\end{figure*}

Since {\lnobi} shares some physical properties with cuprates, it is natural to ask whether the superconducting pairing mechanism and its symmetry also resemble those of the high-{\tc} cuprates. This question has been investigated using different theoretical methods, such as mean-field theory~\cite{gu2023effective,jiang2024high,luo2024high,lu2024interlayer}, cluster dynamical mean-field theory~\cite{Nomura2025}, random phase approximation (RPA)~\cite{zhang2024structural,botzel2024theory,lechermann2023electronic,liu2023s}, fluctuation-exchange approximation~\cite{heier2024competing,sakakibara2024possible}, and functional renormalization group calculations~\cite{yang2023possible,jiang2024theory,zhan2024cooperation}. Unfortunately, different studies yield different symmetries of the superconducting order parameter, even if they use the same numerical method. However, it has been pointed out that the superconducting pairing symmetry of {\lnobi} is highly sensitive, e.g.~to the crystal field splitting between the Ni $e_{g}$ orbitals~\cite{liu2023role}.

Recently, thin films of {\lnobi} have been grown on substrates, so that superconductivity could be observed at ambient pressure~\cite{ko2025signatures,zhou2025ambient}. While there is finally hope to observe the superconducting gap in this material directly, the material is still heavily debated. For example, it remains open whether the $\gamma$ Fermi surface actually exists~\cite{li2025angle,wang2025electronic}. The presence or absence of nodes in the superconducting gap is equally controversial~\cite{shen2025anomalous,sun2025observation,fan2025superconducting}. These results also indicate that superconductivity of {\lnobi} is very sensitive to the specific experimental conditions.

While previous studies often retain only short range hoppings and use strongly downfolded few-orbital models, which do not capture sufficient details of the electronic structure, we use a fully three-dimensional multi-orbital tight-binding model with all relevant states in a wide energy window around the Fermi level that also captures the strong $p$-$d$ hybridization. Our findings show that {\lnobi} hosts spin fluctuations of comparable strength with multiple (incommensurate) wave vectors, which leads to a competition of possible pairing states and probably also explains the absence of magnetic order in this material. Our approach provides a framework for understanding the contradictory results reported in previous studies.

\section{Methods}
We perform density functional theory (DFT) calculations using the full potential local orbital (FPLO) basis set~\cite{Koepernik1999} and the generalized gradient approximation (GGA) to the exchange and correlation potential~\cite{Perdew1996}. Our calculations are based on high-pressure experimental crystal structures of {\lnobi}~\cite{Wang2024} in space group $I4/mmm$ (No.~139), for which we optimized the internal atomic positions within DFT.

Using projective Wannier functions as implemented in FPLO~\cite{Eschrig2009}, we obtain tight-binding models that include all 31 orbitals with \ce{Ni} $3d$ and \ce{O} $2p$ character. Our models almost perfectly reproduce the DFT band energies and Fermi surface, since we retain small-valued and long-range transfer integrals. Keeping the oxygen $2p$ states ensures that relative orbital weights near the Fermi level are correct, which is important when calculating the spin susceptibility at a later stage. For a comparison of DFT, our model and minimal models, see Ref.~\cite{Supplement}.

The kinetic part of the Hamiltonian can be written as:
\begin{equation}
H_{0} = -\sum_{i,j,\sigma} t_{ij}^{sp} c^{\dagger}_{is\sigma} c_{jp\sigma} \,,
\label{eq:HTB}
\end{equation}
where $t_{ij}^{sp}$ are the transfer integrals between sites $i$ and $j$, while $s$ and $p$ are orbital indices, and $\sigma$ is the spin index. Based on the tight-binding models, we calculate the non-interacting susceptibilities $\chi^{pq}_{st}({\bf q})$~\cite{Graser2009}.

We investigate the superconducting pairing from spin fluctuations in {\lnobi} by adding a multi-orbital Hubbard interaction to our kinetic Hamiltonian~\cite{Graser2009}:
\begin{equation}
  \begin{split}
	  H =& \, H_0 
	  	+ U \sum_{i,s} n_{is\uparrow} n_{is\downarrow}	+ \frac{U'}{2} \sum_{i,s,p{\neq}s} n_{is} n_{ip}\\
	  &
		- \frac{J}{2} \sum_{i,s,p{\neq}s} {\bm S}_{is} \cdot
		 {\bm S}_{ip} + \frac{J'}{2} \sum_{i,s,p{\neq}s, \sigma}
		 c^{\dagger}_{is\sigma} c^{\dagger}_{is\bar{\sigma}}
		 c_{ip\bar{\sigma}} c_{ip\sigma} \,.
  \end{split}
  \label{eq:hamiltonian_parameters}
\end{equation}
Here, $c^{\dag}_{is\sigma}$ $(c_{is\sigma})$ denote fermionic creation (annihilation) operators, ${\bm S}_{is}$ is the spin operator and $n_{is\sigma}=c^{\dag}_{is\sigma} c_{is\sigma}$ is the particle number operator. The interaction parameters are intra-orbital Coulomb repulsion $U$, inter-orbital Coulomb repulsion $U'$, Hund's rule coupling $J$, and pair hopping $J'$, which are applied to the Ni $3d$ orbitals and follow the Hubbard-Kanamori relation with $U'=U-2J$ and $J'=J$~\cite{Kanamori1963, Georges2013}.

We employ the multi-orbital random phase approximation~\cite{Graser2009, Altmeyer2016}, which has delivered reliable results for cuprates~\cite{Roig2022}, iron-based materials~\cite{Suzuki2013, Kreisel2017, Shimizu2018}, and organic superconductors~\cite{Aizawa2012, Guterding2016, Mori2018}, to calculate the spin susceptibility and the symmetry of the superconducting pairing within a linearized Eliashberg approach~\cite{Graser2009}. The solutions to this equation are superconducting gap functions with corresponding eigenvalues. The eigenvalue $\lambda$ is a measure of the superconducting pairing strength. To reduce the computational effort, we calculate susceptibilities and pairing only between Ni $3d$ orbitals. We have previously used this approach for multi-orbital models of other materials~\cite{xu2021theory}.

\begin{figure}[t]
\includegraphics[width=\columnwidth]{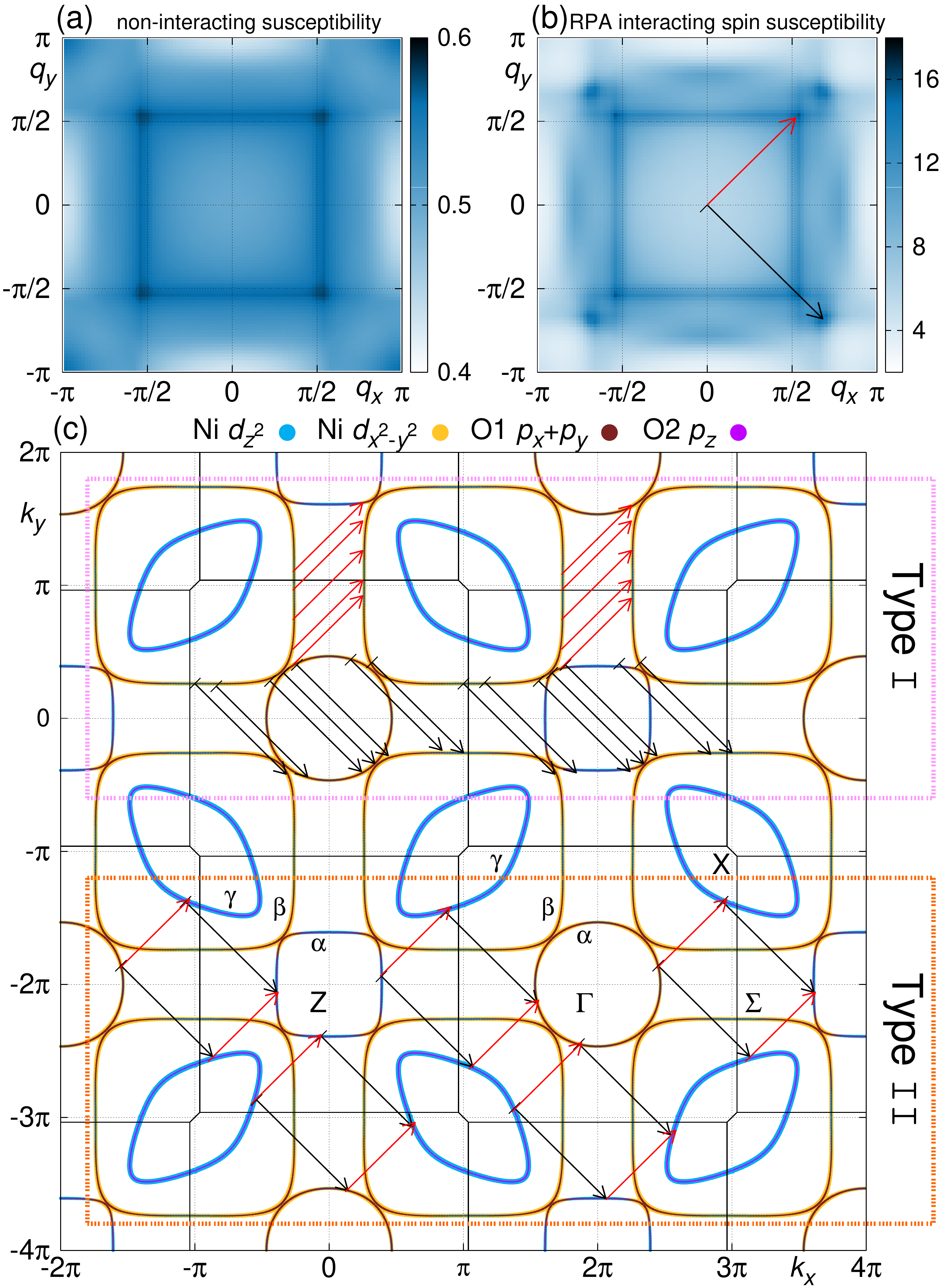}
\caption{Two-dimensional susceptibilities and Fermi surface with nesting vectors for {\lnobi} at a pressure of $P=24.6$~GPa. (a) Non-interacting susceptibility $\chi_0$. (b) RPA interacting spin susceptibility $\chi_s$ for $U=3$~eV and $J=0.75$~eV. The spin fluctuation vector ${\bf q}_1=(\pi/2,\pi/2)$ is shown in red, while ${\bf q}_2\sim(7\pi/10,7\pi/10)$ is shown in black. (c) Two-dimensional cut of the Fermi surface beyond the first Brillouin zone with orbital weights. The \textit{Type I} vectors shown in the top part of the figure are typical nesting vectors, which are related to spin fluctuations. The \textit{Type II} arrangement of spin fluctuations vectors, shown in the lower part of the figure, connects the $\alpha$ and $\gamma$ Fermi surface sheets in a three-dimensional way that is special to space group $I4/mmm$.}
\label{fig:susceptibility_nesting}
\end{figure}

\begin{figure*}[t]
\includegraphics[width=0.95\textwidth]{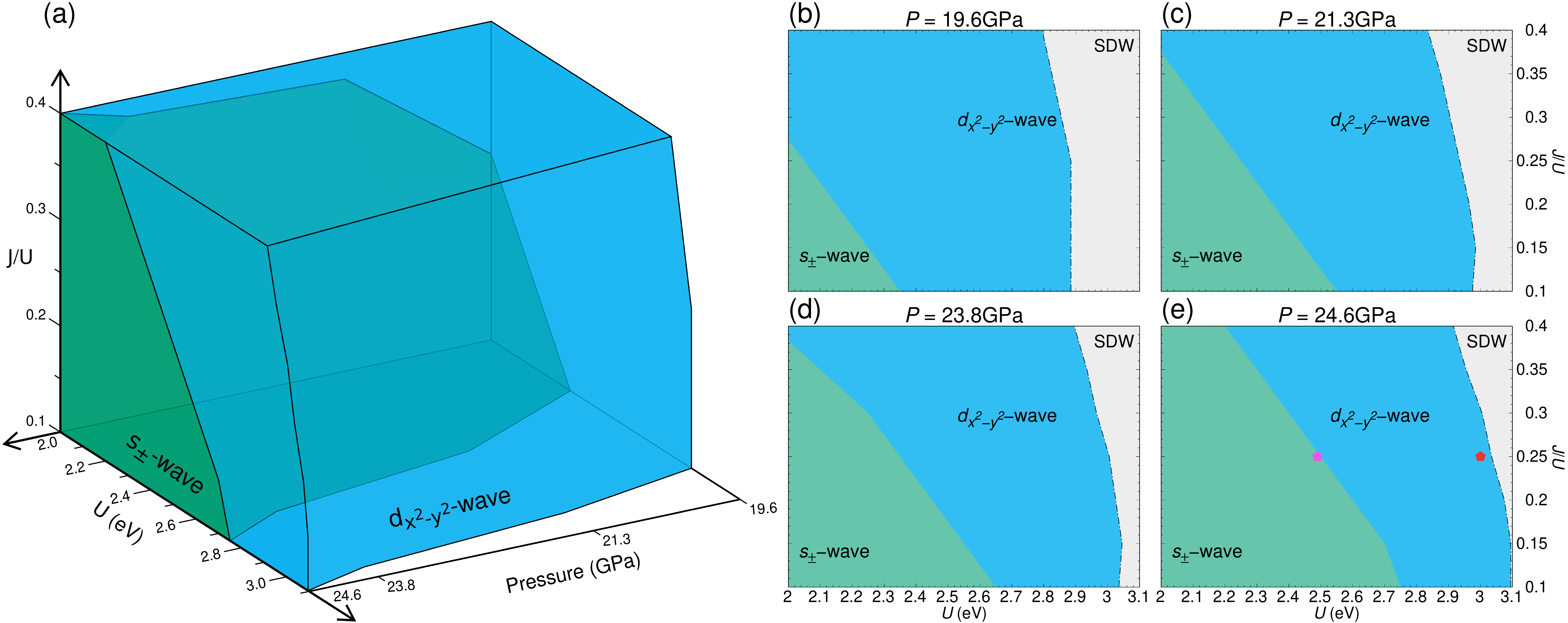}
\caption{Phase diagram of the superconducting pairing symmetry of {\lnobi} as a function of pressure and interaction parameters $U$ and $J$. (a) Three-dimensional phase diagram for sign-changing $s$-wave (green) and $d_{x^2-y^2}$-wave (blue) superconducting gap. (b-e) Two-dimensional phase diagram for each pressure. The pentagons in panel (e) mark the parameter combinations for typical sign-changing $s$-wave and $d_{x^2-y^2}$-wave superconducting gap functions. These are visualized in Fig.~\ref{fig:pairing_lambda_tc}\,(a) and (b).}
\label{fig:sc_phase}
\end{figure*}

\section{Results and Discussion}
We start by investigating the crystal structure and electronic properties of {\lnobi}. In Fig.~\ref{fig:crystal_electronic}\,(a)  we show the crystal structure of {\lnobi} with corner sharing \ce{NiO6} octahedra, and label different atoms with different Wyckoff positions to identify their contributions separately. The oxygen orbitals connect not only the \ce{Ni}-\ce{O} plane, but also the bilayered \ce{Ni} structure with the rest of the system. Due to the bilayered structure, oxygen plays a more important role in nickelates than in cuprates, which is also underscored by the strong hybridization between the O and Ni states and the experimental consequences of oxygen deficiency~\cite{taniguchi1995transport,dong2024visualization,ren2024resolving}.

In Fig.~\ref{fig:crystal_electronic}\,(b), (c), and (d), we analyze the electronic band structure and density of states (DOS) of {\lnobi}. We identify the \ce{Ni} $3d$ orbitals with corresponding weights in Fig.~\ref{fig:crystal_electronic}\,(b) and the relevant orbitals of three different \ce{O} positions in Fig.~\ref{fig:crystal_electronic}\,(d). Importantly, a shallow hole pocket with mostly \ce{Ni} $3d_{z^2}$ character is present in the vicinity of the Fermi level. In the electronic DOS (see Fig.~\ref{fig:crystal_electronic}\,(c)) we find the following largest relative contributions to the DOS at the Fermi level {\nef}: 47.6\% (\ce{Ni} $3d_{z^2}$), 17.6\% (\ce{Ni} $3d_{x^2-y^2}$), 11.3\% (\ce{O}$1$ $2p_{x}+2p_{y}$), and 10.8\% (\ce{O}$2$ $2p_{z}$). These numbers show that discarding the $2p_{x}$ and $2p_{y}$ orbitals of the \ce{O}$1$ position or the \ce{O}$2$ $2p_{z}$ orbital from a low-energy model can not be justified. We now focus on the contributions from the above \ce{Ni} and \ce{O} orbitals, since the other orbitals each contribute less than 2\%.

The importance of the \ce{O}$1$ $2p_{x}$ and $2p_{y}$ orbitals is also obvious from the crystal structure (see Fig.~\ref{fig:crystal_electronic}\,(a)), since they point directly towards the \ce{Ni} atoms within the plane. The large contribution from the oxygen $2p_{z}$ orbitals on the \ce{O}$2$ position, which are perpendicular to the \ce{Ni}-\ce{O} plane, is somewhat surprising and has not been previously considered. Most importantly, this orbital (see Fig.~\ref{fig:crystal_electronic}\,(d)) together with the \ce{Ni} $3d_{z^2}$ orbital (see Fig.~\ref{fig:crystal_electronic}\,(b)) is responsible for the $\gamma$ hole pocket around the X-point. 

The non-interacting susceptibility $\chi_0$ (see Fig.~\ref{fig:susceptibility_nesting}\,(a)) calculated from our 31 band model is a first indication of the momentum structure of spin fluctuations, which result from nesting of the Fermi surface. Note that we take into account all orbital matrix elements when calculating $\chi_0$ (see Ref.~\cite{Supplement}). We now use RPA for the multi-orbital Hubbard Hamiltonian (see eq.~\ref{eq:hamiltonian_parameters}) to calculate the static spin susceptibility $\chi_s$, i.e.~the spin fluctuations on the Fermi surface. The spin susceptibility for pressure $P=24.6$~GPa and interaction parameters $U=3$~eV and $J=0.75$~eV is shown in Fig.~\ref{fig:susceptibility_nesting}\,(b). 
Based on this, we identify the most important nesting vectors. We observe a square-shaped structure of continuously large spin susceptibility similar to literature results~\cite{Zhang2024d, liu2023role}, with strong peaks at the corners (${\bf q}_1=(\pi/2,\pi/2)$). However, other momentum vectors are enhanced more strongly by interactions. In particular, the emerging peak at the incommensurate wave vector ${\bf q}_2\sim(7\pi/10,7\pi/10)$ is directly controlled by the value of $J$, i.e.~the strength of Hund's rule coupling and pair hopping (see Ref.~\cite{Supplement}). The spin fluctuations at ${\bf q}_1\sim(\pi/2,\pi/2)$ and the incommensurate ${\bf q}_2\sim(7\pi/10,7\pi/10)$ are incompatible when it comes to magnetic ordering. This signals a certain magnetic frustration in {\lnobi}, which has previously been identified as an important property of superconducting iron chalcogenides~\cite{glasbrenner2015effect}.

In Fig.~\ref{fig:susceptibility_nesting}\,(c) we show the orbital weights on the Fermi surface in the $k_x$-$k_y$-plane at $k_{z}=0$. The octangle around $\Gamma = (0,0)$ is a two-dimensional cut of the first Brillouin zone. The adjacent squares are two-dimensional cuts of the second Brillouin zone, which in space group $I4/mmm$ are identical to the first Brillouin zone at $k_{z}=\pi$. We identify nesting vectors (\textit{Type I}), which contribute to spin fluctuations in {\lnobi}. Moreover, we find that the spin fluctuation vectors can be arranged in a closed rectangle (\textit{Type II}), which connects the $\gamma$ Fermi surfaces with the central $\alpha$ Fermi surface around both $\Gamma$ and $Z$. This shows that spin fluctuations with both ${\bf q}_1\sim(\pi/2,\pi/2)$ and ${\bf q}_2\sim(7\pi/10,7\pi/10)$ may cooperate when it comes to superconducting pairing.

Previous RPA studies have mostly reported a superconducting gap function of sign-changing ($s_{\pm}$) type~\cite{zhang2024structural,botzel2024theory}, while a minority reported $d$-wave pairing~\cite{lechermann2023electronic}. The mechanism behind the appearance of both order parameters in calculations was partially explained by tuning the crystal field splitting~\cite{liu2023role}. However, all previous studies have used few-orbital models, which also mostly ignored the variation of the electronic structure in the $k_z$ direction and require the use of unrealistically small interaction parameters $U$ and $J$. For an estimate of these parameters from constrained RPA, see Ref.~\cite{christiansson2023correlated}.

These discrepancies appear in few-orbital models, because the weight of missing orbitals w.r.t.~the DFT electronic structure is arbitrarily replaced by the few orbitals included in the low-energy model. This leads to the underestimation of critical interaction values for the \ce{Ni} $3d$ electrons (see also Ref.~\cite{Supplement}) and alters the momentum structure of the resulting susceptibilities due to arbitrary momentum-dependent changes in orbital matrix elements, which inevitably also affect the results of pairing calculations.

We now investigate the superconducting pairing of {\lnobi} using realistic three-dimensional multi-orbital models based on crystal structures at four different pressures~\cite{Wang2024} within RPA. We calculate the symmetry of the leading superconducting pairing instability as a function of pressure, intra-orbital Coulomb interaction $U$ and Hund's rule coupling $J$. Our results in Fig.~\ref{fig:sc_phase}\,(a) show that the superconducting pairing symmetry is sensitive to both pressure and the relative size of interaction parameters. Our phase diagram contains a previously overlooked transition between $s_\pm$- and $d$-wave order parameters, which results from the competition of spin fluctuations with different wave vectors. Both modifications of the electronic structure and changes in interaction parameters tune this competition, which gives rise to different leading symmetries of the superconducting pairing. Therefore, we believe that previous studies correspond to specific points in parameter space, which are incidentally located on different sides of the phase transition between the two pairing symmetries revealed in our study.

As an example, we discuss the superconducting pairing symmetry for the $P=24.6$~GPa case (Fig.~\ref{fig:sc_phase}\,(e)). For small interaction values, the sign-changing $s$-wave ($s_{\pm}$) pairing leads, while increasing $U$ changes the dominant pairing to $d_{x^2-y^2}$ symmetry with opposite phase on the $\alpha$ and $\gamma$ Fermi surfaces (due to the \textit{Type II} arrangement shown in Fig.~\ref{fig:susceptibility_nesting}\,(c)). Increasing the Hund's rule coupling also changes the order parameter from $s_{\pm}$ to $d_{x^2-y^2}$. Our phase diagram also contains areas, in which the spin susceptibility diverges and the RPA becomes unstable. We identify these regions with potential spin-density wave states, denoted as SDW in our phase diagram. With increasing pressure, these regions are shifted to larger interaction values, i.e.~pressure suppresses magnetic ordering tendencies in {\lnobi}. This observation is also consistent with the peculiar shape of the superconducting dome~\cite{li2024pressure}, where high {\tc} is located on the low-pressure side of the dome. Increasing pressure also enlarges the area of the phase diagram covered by the $s_{\pm}$ state, i.e.~the $d_{x^2-y^2}$-wave pairing phase is shifted to higher values of the intra-orbital Coulomb interaction $U$. We also observe that pressure directly increases the crystal field splitting between \ce{Ni} $3d$ states, which leads to an enhancement of $s_{\pm}$-wave pairing. In this sense, our results agree with the findings of Ref.~\cite{liu2023role}.

\begin{figure}[t]
\includegraphics[width=\columnwidth]{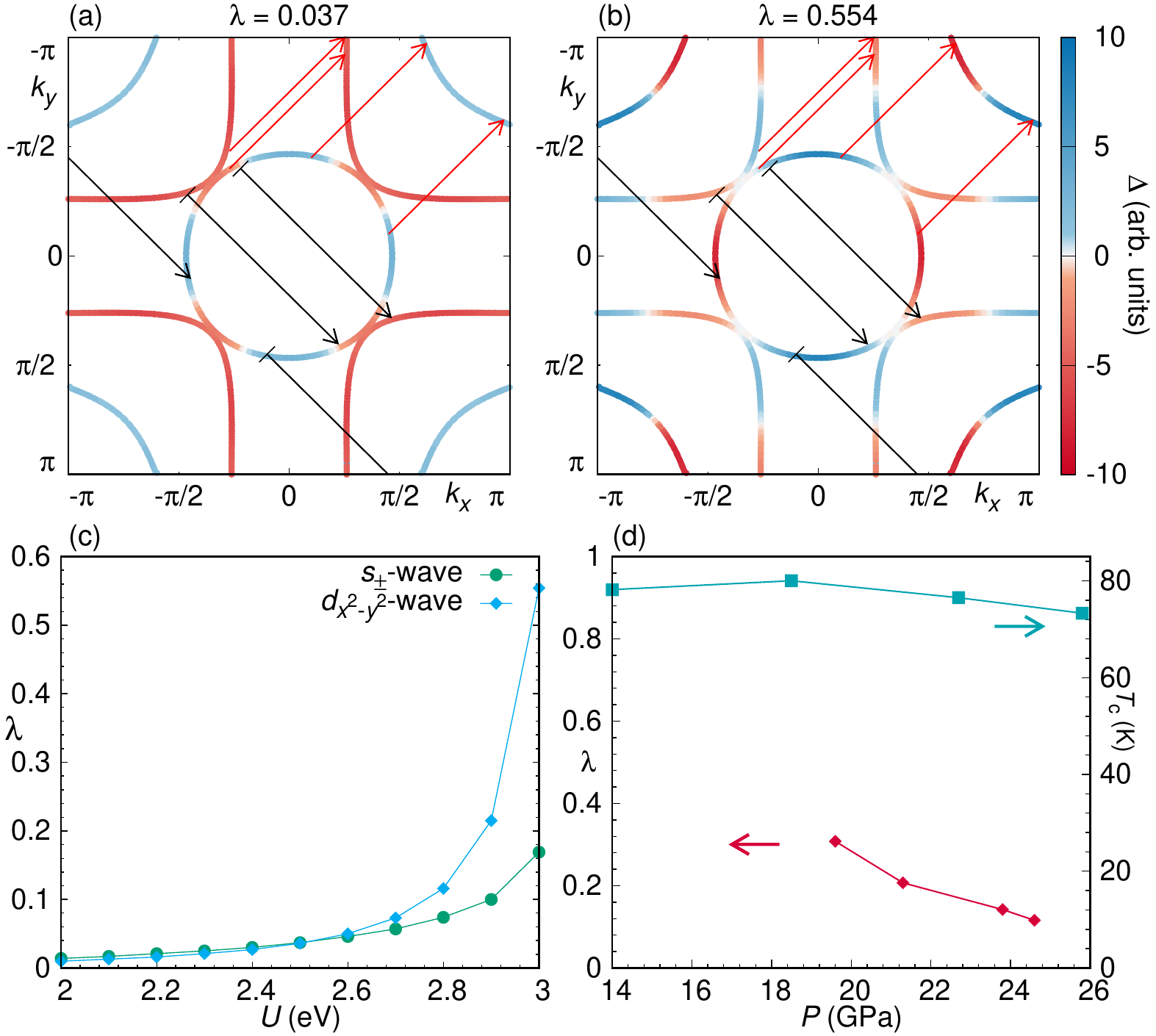}
\caption{(a) Typical sign-changing $s$-wave superconducting gap ($U=2.5$~eV, $J=0.625$~eV and $P=24.6$~GPa). (b) Typical $d_{x^2-y^2}$-wave superconducting gap ($U=3$~eV, $J=0.75$~eV and $P=24.6$~GPa). (c) Competition of eigenvalues $\lambda$ of superconducting gap function for sign-changing $s$-wave (green) and $d_{x^2-y^2}$-wave (blue), where $J=U/4$ at $P=24.6$~GPa. (d) Leading eigenvalue $\lambda$ of the superconducting gap function as a function of pressure at $U=2.8$~eV, $J=0.7$~eV. {$T_{\rm c}^{\rm onset}$} data are from Ref.~\cite{sun2023signatures}.}
\label{fig:pairing_lambda_tc}
\end{figure}

In Fig.~\ref{fig:pairing_lambda_tc}\,(a) and (b), we show examples of $s_{\pm}$- and $d_{x^2-y^2}$-wave gap functions on the Fermi surface at $P=24.6$~GPa. 
The $d_{x^2-y^2}$-wave state contains additional sign changes due to the presence of sub-leading nesting vectors. From the sign-changes of the superconducting gap it is clear that the $d_{x^2-y^2}$-wave state is driven by both the incommensurate ${\bf q}_2\sim(7\pi/10,7\pi/10)$ nesting vector and ${\bf q}_1=(\pi/2,\pi/2)$, where the latter is responsible for the additional sign-change.  The sign-changing $s$-wave state is mostly driven by ${\bf q}_1=(\pi/2,\pi/2)$. Interestingly, the largest absolute values for the gap in the leading $d_{x^2-y^2}$-wave state are located on the shallow $\gamma$ hole pocket. This is not entirely surprising, since the gap scales with the inverse of the Fermi velocity $v_F ({\bf k}) = |\nabla_{\bf k} E({\bf k})|$~\cite{Graser2009}, which is particularly small on $\gamma$.

The competition of $s_{\pm}$- and $d_{x^2-y^2}$-wave is also shown in Fig.~\ref{fig:pairing_lambda_tc}\,(c) with $J=U/4$ at $P=24.6$~GPa. We exhibit the eigenvalue $\lambda$ of the linearized Eliashberg equation for $s_{\pm}$- and $d_{x^2-y^2}$-wave respectively. For small $U$, all eigenvalues are tiny. The eigenvalue of the $s$-wave state is slightly larger than the eigenvalue of the $d_{x^2-y^2}$-wave state. With increasing $U$, $d_{x^2-y^2}$-wave wins the competition gradually and is undoubtedly dominant at $U=3$~eV, which is close to the RPA instability. When we fix the interaction values and plot the leading eigenvalue $\lambda$, which corresponds to the pairing strength, as a function of pressure, we obtain Fig.~\ref{fig:pairing_lambda_tc}\,(d), which shows a moderate decrease, which is reminiscent of the high-pressure region ($P > 18$~GPa) of the superconducting dome in {\lnobi}~\cite{sun2023signatures,li2024pressure}.

\section{Conclusions}
We have shown that the symmetry of the superconducting gap in multi-orbital Hubbard models for {\lnobi} is sensitive to pressure, the value of Coulomb interaction $U$ and the Hund's rule coupling $J$. Pressure directly tunes the crystal field splitting between the Ni $e_{g}$ orbitals. In addition, the increased $p$-$d$ hybridization under pressure leads to increased bandwidths and Fermi velocities, which also negatively affect superconducting pairing, both in agreement with previous reports~\cite{liu2023role, li2024pressure}.

Our results advance the understanding of superconductivity in {\lnobi} in several ways previously not considered: i) Incommensurate spin fluctuations are important and only appear in models with a realistic electronic structure. ii) The strength of incommensurate spin fluctuations is controlled by $J$, i.e.~the Hund's rule and/or pair hopping terms. (The relevance of the pair hopping term is also shown in Ref.~\cite{Nomura2025}.) iii) Spin fluctuations with two different wave vectors cooperate to form a $d_{x^2-y^2}$-wave state for realistic parameter sets, which crucially involves the $\gamma$ hole pocket with small Fermi velocity. On the other hand, these spin fluctuations compete when it comes to magnetic ordering, which indeed appears to be absent in {\lnobi}.

We have shown that our multi-orbital model includes $s_{\pm}$- as well as $d_{x^2-y^2}$-wave pairing states depending on the electronic structure and interaction parameters. Therefore, our results provide a unifying framework to understand the seemingly inconsistent findings based on few-orbital models, which have been reported in the literature. Due to the strong $p$-$d$ hybridization, low-energy models for {\lnobi} should at least include the \ce{Ni} $3d_{z^2}$, \ce{Ni} $3d_{x^2-y^2}$, \ce{O}$1$ $2p_{x}+2p_{y}$ and \ce{O}$2$ $2p_{z}$ states.

Based on our findings, it seems possible that {\lnobi} can be tuned from a $d_{x^2-y^2}$-wave state towards a sign-changing $s$-wave state by applying pressure or another technique that can manipulate the crystal field splitting between Ni $e_{g}$ orbitals and/or the strength of interactions. Due to the importance of the shallow $\gamma$ hole pocket for $d_{x^2-y^2}$-wave pairing in {\lnobi}, we expect that superconductivity in this material is also sensitive to charge doping. This situation is reminiscent of certain iron-based materials, where superconductivity is realized in the vicinity of a Lifshitz transition~\cite{Liu2010, Shi2017, Ren2017, Shimizu2018}.

\acknowledgments
We acknowledge fruitful discussions with Ningning Wang on experiments, and with Harald O. Jeschke, Junya Otsuki, Kazuhiko Kuroki, and Nayuta Takemori on theoretical aspects of this work. We are grateful to Zhijun Wang, Yue Xie, Jingyu Yao, and Ruihan Zhang for enlightening questions and discussions. We are thankful for useful discussions with Xianxin Wu, Hanghui Chen, Qiang-Hua Wang, Meng Wang in IOP conferences and lectures, and revealing questions from Changming Yue, Karsten Held, and Masao Ogata in CCMP 2024. We would also like to thank the organizers of the {\it workshop on theory of cross correlations, multipoles, and computational material design}, held at Gotemba, Japan in February 2025 for providing a stimulating environment and the participants, in particular Yusuke Nomura, for insightful discussions. This work was supported by the National Key R\&D Program of China (Grant No.~2024YFA1408400), and the Center for Materials Genome. The open access publication of this work was funded by Technische Hochschule Brandenburg, University of Applied Sciences.

\bibliography{references}

\clearpage
\includepdf[pages=1]{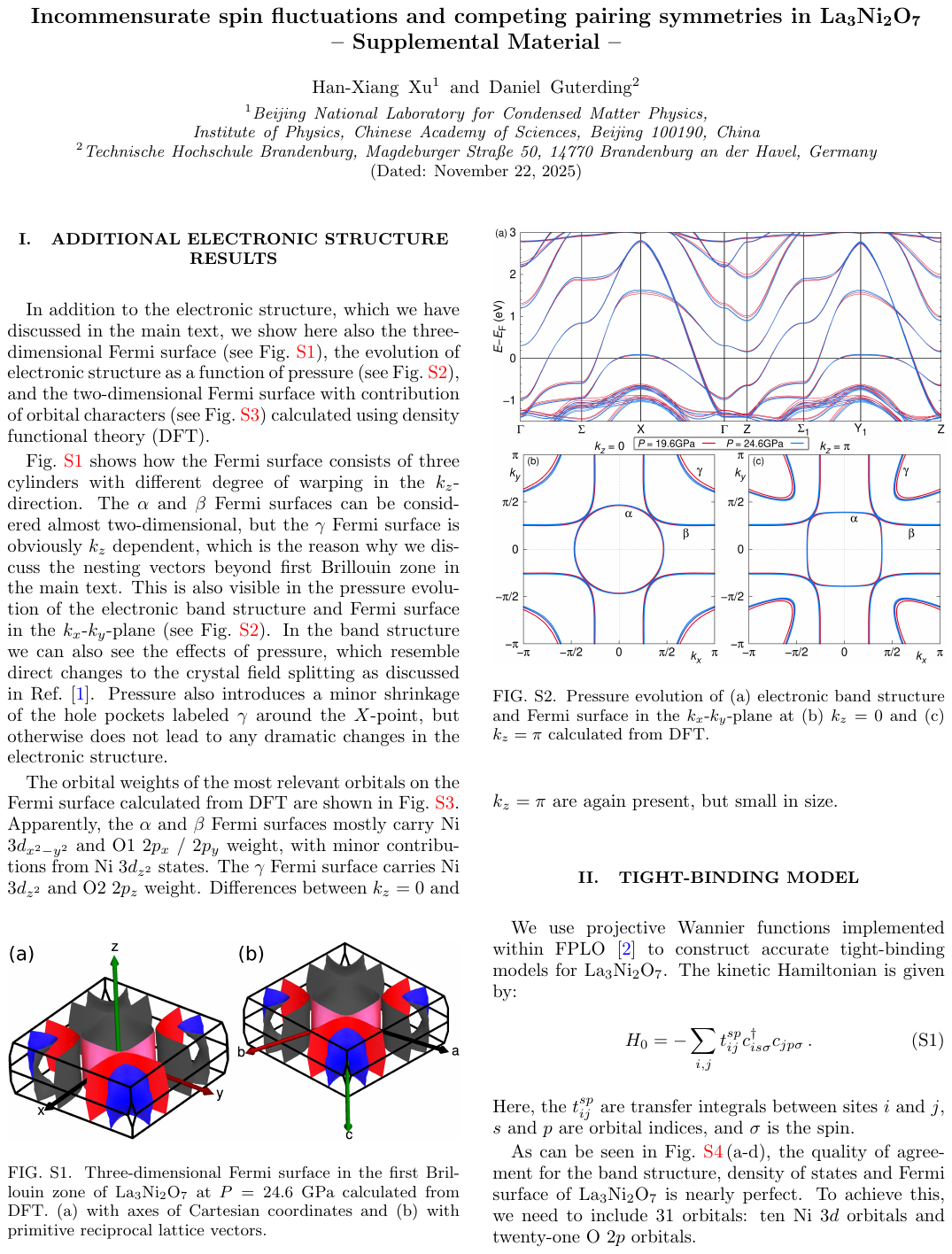}
\clearpage
\includepdf[pages=2]{supplement.pdf}
\clearpage
\includepdf[pages=3]{supplement.pdf}
\clearpage
\includepdf[pages=4]{supplement.pdf}
\clearpage
\includepdf[pages=5]{supplement.pdf}
\clearpage
\includepdf[pages=6]{supplement.pdf}

\end{document}